\documentstyle[12pt]{article}

\def\Ups{\Upsilon}

\def\fun{\mbox{Fun$(G_{q})$}}
\newcommand{\tr}{\triangleright}
\def\rtimes{\mbox{$\times \!\rule{0.3pt}{1.1ex}\,$}}
\def\smash{{\A \rtimes \U}}
\def\R{\mbox{$\cal R$}}

\def\A{\mbox{$\cal A$}}
\def\U{\mbox{$\cal U$}}
\newcommand{\DA}{\Delta _{\cal A}}
\newcommand{\AD}{{}_{\cal A}\Delta }

\def\z{\hspace*{9mm}}
\def\x{\hspace{3mm}}
\newcommand{\ad}{\stackrel{\mbox{\scriptsize ad}}{\triangleright}}
\def\I{\mbox{\boldmath $i$}}
\def\Ix#1{\mbox{\boldmath $i$}_{\chi_#1}}
\def\Li{\hbox{\large\it \pounds}}
\def\Lix#1{\hbox{\large\it \pounds}_{\chi_#1}}
\def\Lio#1#2{\hbox{\large\it \pounds}_{O_#1{}^#2}}
\def\dl{\mbox{\bf d}}

\newcommand{\om}{\mbox{$\omega $}}
\newcommand{\al}{\mbox{$\alpha $}}

\def\tq{\mbox{${\cal T}_q$}}
\def\dg{\mbox{\boldmath$\delta $}}

\def\id{\mathop{\rm id}}
\def\inprod#1#2{
	\left\langle #1, #2\right\rangle}

\begin{document}
\begin{titlepage}
\begin{center}
December 9, 1993     \hfill    LBL-34833\\
                     \hfill    UCB-PTH-93/32\\

\vskip .5in

{\large \bf Cartan Calculus on Quantum Lie Algebras}\footnote{This
work was supported in part by the Director, Office of Energy Research,
Office of High Energy and Nuclear Physics, Division of High Energy
Physics of the U.S. Department of Energy under Contract
DE-AC03-76SF00098 and in part by the National Science Foundation under
grant PHY-90-21139.}

\vskip .5in

Peter Schupp\footnote{SCHUPP@PHYSICS.BERKELEY.EDU}, Paul
Watts\footnote{WATTS@THEORM.LBL.GOV} and Bruno Zumino

\vskip .5in

{\em Department of Physics\\
University of California\\
and\\
Theoretical Physics Group\\
Lawrence Berkeley Laboratory\\
University of California\\
Berkeley, California 94720}
\end{center}
\vskip .5in

\begin{abstract}

A generalization of the differential geometry of forms and vector
fields to the case of quantum Lie algebras is given.  In an abstract
formulation that incorporates many existing examples of differential
geometry on quantum spaces we combine an exterior derivative, inner
derivations, Lie derivatives, forms and functions all into one big
algebra, the ``Cartan Calculus''.

(This is an extended version of a talk presented by P. Schupp at the
XXII$^{th}$ International Conference on Differential Geometric Methods
in Theoretical Physics, Ixtapa, Mexico, September 1993)

\end{abstract}
\end{titlepage}
\newpage
\renewcommand{\thepage}{\arabic{page}}
\setcounter{page}{1}

\section{Introduction}

The central idea behind Connes' Universal Calculus \cite{Connes} in
the context of noncommutative geometry was to retain the classical
differential geometric properties of $\dl$, $i.e.$ nilpotency and the
undeformed Leibniz rule\footnote{We use parentheses to delimit
operations like \dl, $\I_{x}$ and $\Li_{x}$, {\em e.g.} $\dl a =
\dl(a) + a \dl$.  However, if the limit of the operation is clear from
the context, we will suppress the parentheses, {\em e.g.} $\dl(\I_{x}
\dl a) \equiv \dl(\I_{x}(\dl(a)))$.}:  $\dl \alpha = \dl(\alpha ) +
(-1)^{p} \alpha \dl$ for any $p$-form $\alpha $.  Here we want to base
the construction of a differential calculus on quantum groups on two
additional classical formulas: to extend the definition of a Lie
derivative from functions and vector fields to forms we postulate
\begin{equation}
\Li \circ \dl = \dl \circ \Li; \label{LIE}
\end{equation}
this is essential for a geometrical interpretation of vector fields.
The second formula that we can --- somewhat surprisingly --- keep
undeformed in the quantum case is
\begin{equation}
\Lix{i} = \Ix{i} \dl + \dl \Ix{i},\z\mbox{\it (Cartan identity)}
\label{CARTAN}
\end{equation}
where $\{\chi _{i}\}$ are the generators of some quantum Lie algebra.

\section{Quantum Lie Algebras}

A quantum Lie algebra is a Hopf algebra \U\ with a finite-dimensional
biinvariant subvector space $\tq$ spanned by generators $\{\chi_{i}\}$
with coproduct
\begin{equation}
\Delta  \chi _{i} = \chi _{i} \otimes 1 + O_{i}{}^{j} \otimes \chi
_{j}.\label{nicecop}
\end{equation}
More precisely we will call this a quantum Lie algebra of {\bf type
II}.  Let $\{\omega ^{j} \in \tq^{*}\}$ be a dual basis of 1-forms
corresponding to a set of functions $b^{j} \in \A$ via $\omega ^{j}
\equiv S b^{j}_{(1)} \dl b^{j}_{(2)}$; $i.e.$
\begin{eqnarray}
\AD(\chi _{i}) &=& 1\otimes \chi _{i}, \\
\DA(\chi _{i}) &=& \chi _{j} \otimes T^{j}{}_{i},\x T^{j}{}_{i}\in\fun,\\
\I_{\chi _{i}}(\omega ^{j}) &  = & -\inprod{\chi _{i}}{S b^{j}}
= \delta^{j}_{i},\label{ebdual}\\
\AD(\omega ^{i}) &=& 1\otimes \omega ^{i},\\
\DA(\omega ^{i}) &=& \omega ^{j} \otimes S^{-1} T^{i}{}_{j}.
\end{eqnarray}
If the functions $b^{i}$ also close under adjoint coaction $\Delta
^{Ad}(b^{i}) = b^{j}\otimes S^{-1} T^{i}{}_{j}$, we will call the
corresponding quantum Lie algebra one of {\bf type I}.

We can derive two alternate expressions for the exterior derivative of
a function from the Cartan identity (\ref{CARTAN}) in terms of these
bases:
\begin{equation}
\dl(f) = \omega ^{j} \Lix{j}(f) = - \Li_{S \chi _{j}}(f) \omega ^{j}.
\end{equation}
Combining the two expressions for $\dl$ one easily derives the
well-known $f-\omega $ commutation relations
\begin{equation}
f \omega ^{i} = \omega ^{j} \Lio{j}{i}(f).
\end{equation}
The classical limit is given by $O_{j}{}^{i} \rightarrow 1 \delta
^{i}_{j}$, so that forms commute with functions.

\section{Generators, Metrics and the Pure Braid Group}

How does one go about finding the basis of generators $\{\chi _{i}\}$
and the set of functions $\{b^{i}\}$ that define the basis of 1-forms
$\{\omega ^{i}\}$?  Here we would like to present a method that
utilizes pure braid group elements as introduced in \cite{SWZ1}.

Let us recall that a pure braid element $\Ups$ is an element of
$\U\hat{\otimes}\U$ that commutes with all coproducts of elements of
\U, $i.e.$
\begin{equation}
\Ups \Delta (y) = \Delta (y) \Ups,\z \forall y \in \U.
\end{equation}
$\Ups$ maps elements of \A\ to elements of \U\ with special
transformation properties under the right coaction:
\begin{equation}
\begin{array}{l}
\Ups:\A \to  \U\,:\x b \mapsto \Ups_{b} \equiv \inprod{\Ups}{b \otimes
\id};\\
\DA(\Ups_{b}) = \Ups_{b_{(2)}} \otimes S(b_{(1)}) b_{(3)} = \inprod{\Ups
\otimes \id}{\tau ^{23}(\Delta ^{Ad}(b) \otimes \id)}.
\end{array}
\end{equation}
An element $\Ups$ of the pure braid group defines furthermore a
bi\-linear qua\-dra\-tic form on \A\,
\begin{equation}
(\:,\:): \A \otimes \A \to  k:\x a \otimes b \mapsto (a,b) =
-\inprod{\Ups}{a\otimes S(b)} \in k,
\end{equation}
with respect to which we can construct orthonormal bases $\{b_{i}\}$
and $\{b^{j}\}$ of functions ($i.e.$ $(b_{i},b^{j}) = \delta
_{i}^{j}$) that in turn will give generators $\chi _{i} :=
\Ups_{b_{i}}$ and 1-forms $\omega ^{j} := S(b^{j}_{(1)}) \dl b^{j}_{(2)}$.
Typically, one can choose span$\{b_{i}\}$ = span$\{b^{j}\}$; then one
starts by constructing one set, say $\{b_{i}\}$, of functions that
close under adjoint coaction
\begin{equation}
\Delta ^{Ad} b_{i} = b_{j} \otimes T^{j}{}_{i}.
\end{equation}
If the numerical matrix
\begin{equation}
\eta _{ij} := -\inprod{\Ups}{b_{i} \otimes Sb_{j}}\z\mbox{\it (metric)}
\end{equation}
is invertible, $i.e.$ det$(\eta ) \neq 0$, then we can use its inverse
$\eta ^{ij} := (\eta ^{-1})_{ij}$ to raise indices via
\begin{equation}
b^{i} = b_{j} \eta ^{ji}.
\end{equation}
This metric is invariant --- or $T$ is orthogonal --- in the sense
that
\begin{equation}
\eta _{ji}  = \eta _{kl} T^{k}{}_{j} T^{l}{}_{i}.
\end{equation}
Once we have obtained a metric $\eta $, we can truncate the pure braid
element $\Ups$ and work instead with
\begin{equation}
\Ups \to  \Ups_{trunc} =-S(\chi _{i}) \otimes \chi ^{i} = -S(\chi_{i})
\otimes \chi _{j} \eta ^{ji},
\end{equation}
which also commutes with all coproducts.  Casimir operators can also
be constructed from elements of the pure braid group. The truncated
pure braid element gives, for instance, the quadratic casimir
\begin{equation}
[\cdot  \circ \tau  \circ (S^{-1} \otimes \id)](\Ups_{trunc}) = \eta
^{ji} \chi _{j} \chi _{i}.\z\mbox{\it (casimir)}
\end{equation}
Now we would like to show that we have actually obtained a quantum Lie
algebra of type I\footnote{Note, that $\Ups$ has to be carefully
chosen to insure the correct number of generators. Furthermore, we
still have to check the coproduct of the generators. If they are not
of the form $\Delta \chi _{i} = \chi _{i} \otimes 1 + O_{i}{}^{j}
\otimes \chi _{j}$ then we might still consider a calculus with
deformed Leibniz rule.}:
\begin{equation}
-\inprod{\chi _{i}}{S b^{j}} = -\inprod{\Ups}{b_{i} \otimes Sb_{k}}
\eta ^{kj} = \eta _{ik}
\eta ^{kj} = \delta _{i}^{j},
\end{equation}
\begin{equation}
\DA(\chi _{i}) = \Ups_{b_{i(2)}} \otimes S(b_{i(1)}) b_{i(3)} =
\Ups_{b_{j}} \otimes T^{j}{}_{i} =\chi _{j} \otimes T^{j}{}_{i}
\end{equation}
and
\begin{equation}
\Delta ^{Ad}(b^{i})
         =  b_{k} \otimes T^{k}{}_{j} \eta ^{ji}
         =  b_{k} \otimes \eta ^{kl} \eta _{ln} T^{n}{}_{j} \eta ^{ji}
         =  b^{k} \otimes S^{-1} T^{i}{}_{k}.
\end{equation}

\subsection{Examples}

\subsubsection{The $R$-matrix approach}

Often one can take $b_{i} \in $span$\{t^{n}{}_{m}\}$, where
$t^{n}{}_{m}$ is a quantum matrix in the defining representation of
the quantum group under consideration. If we are dealing with a
quasitriangular Hopf algebra with universal $\R \equiv \alpha _i
\otimes \beta ^i$, a natural choice for the pure braid element is
\begin{equation}
\Ups_{R} = \frac {1}{\lambda }\left(1 \otimes 1 - \R^{21} \R^{12}\right),
\end{equation}
where the term $\R^{21} \R^{12}$ has been introduced and extensively
studied in \cite{RS} and later in \cite{Jur,Maj,SWZ1}.  These choices
of $b_{i}$s and $\Ups$ lead to the $R$-matrix approach to differential
geometry on quantum groups.  The metric is
\begin{equation}
\eta  = -\inprod{X_{1}}{St_{2}} = \frac {1}{\lambda
}\left(\left[\left(R_{21}{}^{-1}\right)^{t_{2}}
\left(R_{12}{}^{t_{2}}\right)^{-1}\right]^{t_{2}} - I \right),
\end{equation}
where $X_{1} = \inprod{\Ups_{R}}{t_{1} \otimes \id}$, $R_{12}=
\inprod{\R}{t_{1}\otimes t_{2}}$, and $I$ is the identity matrix.  In
the case of GL${}_{q}(2)$ we find
\begin{equation}
\eta _{\mbox{\tiny GL${}_{q}(2)$}} = \left(
        \begin{array}{cccc}
                -q^{-3} & 0 & 0 & 0 \\
                0 & 0 & -q^{-1} & 0 \\
                0 & -q^{-3} & 0 & 0 \\
                0 & 0 & 0 & -q^{-1}
        \end{array}\right).\label{peter}
\end{equation}
Using this metric we recover --- as expected --- the well-known
\cite{Zum,SWZ2} expression of the exterior derivative $\dl$ on
functions in terms of the quantum trace over $X$ and the Cartan-Maurer
form $\Omega = t^{-1} \dl t$:
\begin{equation}
\dl = \omega ^{i} \chi _{i} = - \mbox{tr}(D^{-1} \Omega X)
\z\mbox{\it (on functions)}.
\end{equation}
(This follows essentially from $D^{-1}_{2} \eta _{12} = P_{12}$, where
$D = \inprod{u}{t}$ with $u = S(\beta^i) \alpha_i$ and $P$ is the
permutation matrix.)

\subsubsection{Trace formula for the metric}

Again in the case where $\U$ is a quasitriangular Hopf algebra, there
exists an alternate way of defining a Killing form; let $\rho : \U
\rightarrow M_{n}(k)$ be an $n \times n$ matrix representation of $\U$
with entries in $k$.  Define the map $\eta^{(\rho)}: \U \otimes \U
\rightarrow k$ as\footnote{The map $\eta^{(\rho)}$ as defined in the
Proceedings of the XXII${}^{th}$ DGM Conference differs from the one
appearing here by an antipode. This can be compensated by choosing the
contragredient representation in the former case.}
\begin{equation}
\eta^{(\rho)}(x \otimes y):=\mbox{tr}_{\rho}\left[ S(uxy) \right],
\end{equation}
where $x,y \in \U$, tr$_{\rho}$ is the trace over the given
representation, and $u$ (see above) implements the square of the
antipode \cite{Drin}.  The map $\eta^{(\rho)}$ has the following properties:
\begin{eqnarray}
\eta^{(\rho)}(y\otimes x)&=&\eta^{(\rho)}(x\otimes S^2 (y)),\\
\eta^{(\rho)}((z_{(1)}\ad x)\otimes (z_{(2)}\ad y))&=&
\eta^{(\rho)}(x \otimes y)\epsilon(z),
\end{eqnarray}
for all $x, y, z \in \U$.  Respectively, these express the symmetry of
$\eta^{(\rho)}$ and its invariance under the adjoint action.  In the
case when $\U$ is a quantum Lie algebra with generators $\{
\chi_i \}$, we can define the Killing metric for the representation
$\rho$ as
\begin{equation}
\eta^{(\rho)}_{ij}:=\eta^{(\rho)}(\chi_i \otimes \chi_j ).
\end{equation}
For the quantum group GL$_q(2)$, with $\rho$ being the fundamental
representation, a calculation gives the Killing metric (expressed as a
matrix in the basis $(\chi_1 , \chi_+ , \chi_- , \chi_2 )$) as
\begin{eqnarray}
\eta^{(\mbox{\tiny fund GL$_q (2)$})}&=&\left(
\begin{array}{cccc}
q^{-7} & 0 & 0 & 0\\
0 & 0 & q^{-5} & 0\\
0 & q^{-7} & 0 & 0\\
0 & 0 & 0 & q^{-5}
\end{array}
\right)\nonumber \\
&=&-q^{-4}\eta_{\mbox{\tiny GL$_q (2)$}} , \label{paul}
\end{eqnarray}
so we see that the two differ only by a multiplicative constant (which
is $-q^{-2n}$ for the GL$_q (n)$ case).  If we reduce the two matrices
(\ref{peter}) and (\ref{paul}) into $1 \oplus 3$ matrices,
corresponding to the basis $(\chi_1 + q^{-2}\chi_2 , \chi_+ , \chi_-
,\chi_1 - \chi_2 )$, we find
\begin{equation}
\eta_{\mbox{\tiny GL$_q (2)$}}=-q^4 \eta^{(\mbox{\tiny fund GL$_q
(2)$})}=- \left(
\begin{array}{cccc}
q^{-3}+q^{-5} & 0 & 0 & 0\\
0 & 0 & q^{-1} & 0\\
0 & q^{-3} & 0 & 0\\
0 & 0 & 0 & q^{-1}+q^{-3}
\end{array}\right) .
\end{equation}
Here we see the decomposition which in the undeformed case is
expressed as GL(2)$\simeq$U(1)$\times$SL(2).

Further properties of the Killing metric as defined in this
subsection will be examined in a forthcoming paper \cite{Watts}.

\subsubsection{The 2-dim quantum euclidean group}

This is an example of a quantum Lie algebra that seems to have no
universal $\R$ and where the set of functions $\{b_{i}\}$ does not
arise from the matrix elements of some quantum matrix. In \cite{SWZ1}
we constructed such a set of functions $b_0$, $b_+$, $b_-$, $b_1$, and
a pure braid element $\Ups_{e}= \frac {1}{\lambda }(\Delta c - c
\otimes 1)$ from the casimir $c := P_+ P_-$ of e${}_q(2)$.  Now we can
put the new machinery to work and calculate the (invertible) metric
\begin{equation}
\eta _{\mbox{\tiny E${}_{q}(2)$}} = \left(
        \begin{array}{cccc}
                0 & 1 & 0 & 0 \\
                1 & 0 & 0 & 0 \\
                0 & 0 & 0 & -1 \\
                0 & 0 & -q^{-2} & 0
        \end{array}\right),
\end{equation}
which immediately gives an expression for $\dl$ on functions:
\begin{equation}
\dl = \omega _{0} \chi _{1} + \omega _{1} \chi _{0} - q^{2}\omega _{+}
\chi _{-} - \omega _{-} \chi _{+}.
\end{equation}

\subsubsection{Universal Calculus} \label{S:UC}

Given (countably infinite) linear bases $\{ e_{i}\}$ and $\{f^{i}\}$
of the Hopf algebras \U\ and \A\ respectively, we can always construct
new counit-free elements $e_{i} - 1_{\cal U} \epsilon (e_{i})$ and
$f^{i} - 1_{\cal A} \epsilon (f^{i})$ that each span (infinite)
biinvariant spaces $\tq$ and ${\cal C}_q$ respectively and have
coproducts of the form (\ref{nicecop}); in fact $1_{\cal U} \oplus \tq
= \U$ and $1_{\cal A} \oplus {\cal C}_q = \A$ as vector spaces.  Using
some Gram-Schmitt orthogonalization procedure one can rearrange the
infinite bases of \U\ and \A\ in such a way that $e_{0} = 1_{{\cal
U}}$, $f^{0} = 1_{{\cal A}}$ and $e_{i}$, $f^{i}$ with $\epsilon
(e_{i}) = \epsilon (f^{i}) = 0$ for $i = 1,\ldots,\infty $ span $\tq$
and ${\cal C}_q$ respectively. (In the rest of this section roman
indices $i,j,k,\ldots$ will only take on values from $1$ to $\infty
$.) To avoid confusion with the finite-dimensional quantum Lie
algebras, we will use the symbol $\dg$ instead of $\dl$ for the
exterior derivative.

Given orthonormal linear bases $\{e_{i}\}$ and $\{f^{i}\}$ of $\tq$
and ${\cal C}_q$ we can now express \dg\ on functions $a \in \A$ as
\begin{equation}
\dg (a) = - \om_{S^{-1}f^{i}} \Li_{e_{i}-1 \epsilon (e_{i})}(a);
\label{hopfd}
\end{equation}
note, however, that {\em all} of these $\om_{S^{-1}f^{i}}$s are
treated as linearly independent and remain so even in the classical
limit, because (\ref{hopfd}) in conjunction with the Leibniz rule for
\dg\ only gives trivial commutation relations between forms and
functions ($a\om_{b} =\om_{bS^{-1}a_{(2)}} a_{(1)} - \epsilon (b)
\om_{S^{-1}a_{(2)}} a_{(1)}$); therefore, it is not generally possible
to reorganize the infinite set of $\om_{S^{-1}f^{i}}$s into a finite
basis of 1-forms.  This is the case for Connes' noncommutative
geometry \cite{Connes} and is in contrast to the ordinary ``textbook''
treatment of differential calculi that has forms commuting with
functions.

\section{Calculus of Functions, Vector Fields and Forms}

Here we will generalize the Cartan calculus of ordinary {\em
commutative} differential geometry to the case of quantum Lie
algebras.

As in the classical case, the Lie derivative of a function is given by
the action of the corresponding vector field, {\em i.e.}
\begin{equation}
\begin{array}{l}
\Lix{i}(a) = \chi_i \tr a =  a_{(1)} \inprod{\chi_i}{a_{(2)}},\\
\Lix{i} a = a_{(1)} \inprod{\chi_{i(1)}}{a_{(2)}} \Li_{\chi_{i(2)}}.
\end{array} \label{XA}
\end{equation}
The action on products is given through the coproduct of $\chi_i$:
\begin{equation}
\chi_i \tr a b = (\chi_{i(1)} \tr a)(\chi_{i(2)} \tr b).
\label{XAB}
\end{equation}
The Lie derivative along $\chi_i$ of an element $y \in \U$ is given by
the adjoint action in \U:
\begin{equation}
\Lix{i}(y) = \chi_i \ad y =  \chi_{i(1)} y S(\chi_{i(2)}).
\label{xady}
\end{equation}
To find the action of $\Ix{i}$ we can now use the Cartan identity
(\ref{CARTAN}):
\begin{equation}
\chi_i \tr a  =  \Lix{i}(a)
        =  \Ix{i}(\dl a) + \dl(\Ix{i} a).
\end{equation}
As the inner derivation $\Ix{i}$ contracts 1-forms and is zero on
0-forms like $a$, we find
\begin{equation}
\Ix{i}(\dl a) = \chi _{i} \tr a = a_{(1)} \inprod{\chi _{i}}{a_{(2)}}.
\label{incomplete}
\end{equation}
Next consider that for any form $\al$,
\begin{equation}
\Lix{i}(\dl \al ) =  \dl(\Ix{i} \dl \al ) + \Ix{i}(\dl \dl \al )
=  \dl(\Lix{i} \al ) + 0, \label{Ld}
\end{equation}
which shows that Lie derivatives commute with the exterior derivative;
$\Lix{i} \dl = \dl \Lix{i}$ (we will also assume $\Li_x$ commutes with
$\dl$ for {\em all} $x\in \U$ as well).  From this and (\ref{XA}) we
find
\begin{equation}
\Lix{i} \dl(a) = \dl(a_{(1)}) \inprod{\chi_{i(1)}}{a_{(2)}}
\Li_{\chi_{i(2)}}.
\end{equation}
To find the complete commutation relations of $\Ix{i}$ with functions
and forms rather than just its action on them, we next compute the
action of $\Lix{i}$ on a product of functions $a$, $b$ $\in \A$,
$i.e.$
\begin{equation}
\Lix{i}(a b) =  \Ix{i} \dl(a b) =  \Ix{i}(\dl(a) b + a \dl(b)),
\end{equation}
and compare with equation (\ref{XAB}).  Recalling that the $\chi _{i}$
have coproducts of the form $\Delta \chi _{i} = \chi _{i} \otimes 1 +
O_{i}{}^{j} \otimes \chi _{j}$, $O_{i}{}^{j} \in \U$, we obtain
\begin{equation}
\I_{\chi _{i}} a = (O_{i}{}^{j} \tr a) \;\I_{\chi _{j}}
= \Li_{O_{i}{}^{j}}(a) \;\I_{\chi _{j}} \label{lcross}
\end{equation}
if we assume that the commutation relation of $\I_{\chi _{i}}$ with
$\dl(a)$ is of the general form
\begin{equation}
\I_{\chi _{i}} \dl(a) = \underbrace{\I_{\chi _{i}}(\dl a)}_{\in \A} +
\mbox{``braiding term''}\cdot \I_{\chi _{?}}\,.
\end{equation}
A calculation of $\Li_{\chi _{i}}(\dl(a) \dl(b))$ along similar lines
gives in fact
\begin{equation}
\begin{array}{rcl}
\I_{\chi _{i}} \dl(a) &=& (\chi _{i} \tr a) - \dl(O_{i}{}^{j} \tr a)
\;\I_{\chi _{j}}\\
&=&\I_{\chi _{i}}(\dl a) - \Li_{O_{i}{}^{j}}(\dl a) \;\I_{\chi _{j}},
\end{array}
\end{equation}
and we propose for any $p$-form $\al$:
\begin{equation}
\I_{\chi _{i}} \al = \I_{\chi _{i}}(\al) +  (-1)^{p}
\Li_{O_{i}{}^{j}}(\al)\; \I_{\chi _{j}}.
\end{equation}
Using the Cartan identity we can derive commutation relations for
(Lie) derivatives and functions from equation (\ref{lcross}), which
can be written in Hopf algebra language as
\begin{equation}
\chi_i a = a_{(1)} \inprod{\chi_{i(1)}}{a_{(2)}} \chi_{i(2)}.
\end{equation}
This actually defines the product in the cross-product algebra
$\smash$ of general vector fields that one obtains by combining
the Hopf algebras \A\ and \U; see $e.g.$ \cite{SWZ1}.

\subsection{Maurer-Cartan Forms}

The most general left-invariant 1-form can be written as \cite{Wor}
\begin{equation}
\om_{b} := S(b_{(1)}) \dl(b_{(2)}) = - \dl(S b_{(1)} ) b_{(2)}
\end{equation}
\begin{equation}
(\mbox{\em left-invariance:}\x\AD(\om_{b}) =
S(b_{(2)}) b_{(3)} \otimes S(b_{(1)}) \dl(b_{(4)})
= 1 \otimes \om_{b} ),
\end{equation}
corresponding to a function $b \in \A$. If this function happens to be
$t^{i}{}_{k}$, where $t \in M_{m}(\A)$ is an $m \times m$ matrix
representation of \U\ with $\Delta (t^{i}{}_{k}) =$\mbox{$t^{i}{}_{j}
\otimes t^{j}{}_{k}$} and $S(t)=t^{-1}$, we obtain the well-known
Cartan-Maurer form $\om_{t} = t^{-1} \dl(t)$. Here is a nice formula
for the exterior derivative of $\om_{b}$:
\begin{equation}
\dl(\om_{b})  = - \om_{b_{(1)}} \om_{b_{(2)}}.
\end{equation}
The Lie derivative is
\begin{equation}
\Lix{i}(\om_{b}) = \om_{b_{(2)}} \inprod{\chi_i}{S(b_{(1)})b_{(3)}}.
\label{XOM}
\end{equation}
The contraction of left-invariant forms with $\Ix{{}}$ is
\begin{equation}
\Ix{i}(\om_{b}) =  -\inprod{\chi_i}{S(b)} \in k.
\label{IOM}
\end{equation}

\subsection{Tensor Product Realization of the Wedge}

{}From (\ref{XOM}) and (\ref{IOM}) we find commutation relations for
$\I_{\chi _{i}}$ with $\om^{j}$,
\begin{equation}
\begin{array}{rcl}
\I_{\chi _{i}} \om^{j}
&=& \delta _{i}^{j} - \Li_{O_{i}{}^{k}}(\om^{j}) \I_{\chi _{k}}\\
&=& \delta _{i}^{j} - \om^{m} \inprod{O_{i}{}^{k}}{S^{-1}(T^{j}{}_{m})}
\I_{\chi _{k}},
\end{array} \label{IOMI}
\end{equation}
which can be used to define the wedge product $\wedge$ of forms as
a kind of antisymmetrized tensor product\footnote{So far we have
suppressed the $\wedge$-symbol; to avoid confusion we will reinsert it
in this paragraph.}.  As in the classical case we make an ansatz for
the product of two forms in terms of tensor products
\begin{equation}
\om^{i} \wedge \om^{j} = \om^{i} \otimes \om^{j} - \hat{\sigma }^{ij}{}_{mn}
\om^{m} \otimes \om^{n},
\end{equation}
with as yet unknown numerical constants $\hat{\sigma }^{ij}{}_{mn} \in
k$, and define $\I_{\chi _{i}}$ to act on this product by contracting
in the first tensor product space, {\em i.e.}
\begin{equation}
\I_{\chi _{i}}(\om^{j} \wedge \om^{k}) = \delta _{i}^{j} \om^{k} -
\hat{\sigma }^{jk}{}_{mn} \delta _{i}^{m} \om^{n}.
\end{equation}
But from (\ref{IOMI}) we already know how to compute this, and we find
\begin{equation}
\hat{\sigma }^{ij}{}_{mn} = \inprod{O_{m}{}^{j}}{S^{-1}(T^{i}{}_{n})},
\end{equation}
or
\begin{equation}
\begin{array}{rcl}
\om^{i} \wedge \om^{j}
&=& (I-\hat{\sigma })^{ij}{}_{mn}\om^{m} \otimes \om^{n} \\
&=& \om^{i} \otimes \om^{j} - \om^{k} \otimes \Li_{O_{k}{}^{j}}(\om^{i}).
\end{array}
\end{equation}
These equations give implicit (anti)commutation relations between the
$\om^{i}$s. Note that $(I-\hat{\sigma })$ has a sensible classical
limit --- it becomes $(I-P)$ where $P$ is the permutation matrix.
Using the same method as for $\omega $ we can also obtain a tensor
product decomposition of products of inner derivations.
\paragraph{Example: Maurer-Cartan Equation}
\begin{equation}
\begin{array}{rcl}
\dl \omega ^{j}  & = & \dl \omega _{b^{j}} = -\omega _{b^{j}_{(1)}}
\wedge \omega _{b^{j}_{(2)}}\\
& = & -\omega _{S^{-1}(Sb^{j}_{(1)} b^{j}_{(3)})} \otimes \omega
_{b^{j}_{(2)}}\\
& = & -\omega ^{k} \otimes \omega ^{l} \inprod{-S\chi_{k}}{S^{-1}(Sb^{j}_{(1)}
b^{j}_{(3)})}\inprod{-S\chi _{l}}{b^{j}_{(2)}}\\
& = & -\omega ^{k} \otimes \omega ^{l} \inprod{\underbrace{(S^{-1}\chi
_{k})_{(1)} \chi _{l} S(S^{-1}\chi _{k})_{(2)}}_{S^{-1}\chi _{k} \ad
\chi _{l}}}{Sb^{j}} \\
& = & -f'_{k}{}^{j}{}_{l} \omega ^{k} \otimes \omega ^{l}.
\end{array}
\end{equation}
In the previous equation we have introduced the adjoint action of a
left-invariant vector field on another vector field. A short
calculation gives
\begin{equation}
S^{-1}\chi _{k} \ad \chi _{l}
= \chi _{b} \chi _{c} (\delta ^{c}_{k}\delta ^{b}_{l} - \hat{\sigma
}^{cb}{}_{kl}) = \chi _{a} \inprod{S^{-1}\chi _{k}}{T^{a}{}_{l}}=\chi _{a}
f'{}_{k}{}^{a}{}_{l}
\end{equation}
as compared to
\begin{equation}
\chi _{k} \ad \chi _{l} \equiv \Li_{\chi _{k}}(\chi _{l}) = \chi _{b}
\chi _{c} (\delta ^{c}_{k}\delta ^{b}_{l} - \hat{R}^{cb}{}_{kl})
= \chi _{a} f_{k}{}^{a}{}_{l},
\end{equation}
with $\hat{R}^{cb}{}_{kl} = \inprod{O_{k}{}^{b}}{T^{c}{}_{l}}$.  The
two sets of structure constants are related by $\inprod{\chi
_{k}}{T^{a}{}_{l}}= f_{k}{}^{a}{}_{l} = -f'_{i}{}^{a}{}_{l}
R^{ij}{}_{kl}$.  See
\cite{CaMo} for a detailed discussion of such structure constants.

\subsubsection{The ``Anti-Wedge'' Operator}

There is actually an operator $W$ that recursively translates wedge
products into the tensor product representation:
\begin{equation}
\begin{array}{l}
W: \Lambda ^{p}_{q} \rightarrow
{\cal T}^{*}_{q} \otimes \Lambda ^{p-1}_{q},\x p \geq 1,\\
W(\alpha ) = \om^{n} \otimes \I_{\chi _{n}}(\alpha ),
\end{array}
\end{equation}
for any $p$-form $\alpha $.  Two examples:
\begin{equation}
\begin{array}{rcl}
\om^{j} \wedge \om^{k}
        & = & \om^{n} \otimes \I_{\chi _{n}}(\om^{j} \wedge \om^{k})\\
        & = & \om^{n} \otimes
                (\delta ^{j}_{n} \om^{k} - \Li_{O_{n}{}^{m}}(\om^{j})
                \delta ^{k}_{m})\\
        & = & \om^{j} \otimes \om^{k} - \om^{n} \otimes
                \Li_{O_{n}{}^{k}}(\om^{j})\\
        & = & \om^{j} \otimes \om^{k} - \om^{n} \otimes \om^{m}
                \hat{\sigma }^{jk}{}_{nm}
\end{array}
\end{equation}
and, after a little longer computation that uses $W$ twice,
\begin{equation}
\begin{array}{rcl}
\om^{a} \wedge \om^{b} \wedge \om^{c}
& = & \om^{a} \otimes (\om^{b} \wedge \om^{c}) - \om^{i} \otimes
(\om^{j} \wedge \om^{c}) \hat{\sigma }^{ab}{}_{ij}\\
&  & + \om^{i} \otimes (\om^{j} \wedge \om^{k}) \hat{\sigma }^{al}{}_{ij}
                \hat{\sigma }^{bc}{}_{lk}\\
& = & \om^{a} \otimes \om^{b} \otimes \om^{c}
  - \om^{a} \otimes \om^{j} \otimes \om^{k} \hat{\sigma }^{bc}{}_{jk}\\
 &   & -\om^{i} \otimes \om^{j} \otimes \om^{c} \hat{\sigma }^{ab}{}_{ij}
                + \om^{i} \otimes \om^{j} \otimes \om^{k}
                \hat{\sigma }^{lc}{}_{jk} \hat{\sigma }^{ab}{}_{il}\\
 &   & + \om^{i} \otimes \om^{j} \otimes \om^{k}
         \hat{\sigma }^{al}{}_{ij} \hat{\sigma }^{bc}{}_{lk}
         - \om^{i} \otimes \om^{j} \otimes \om^{k}\hat{\sigma
}^{an}{}_{il} \hat{\sigma }^{bc}{}_{nm} \hat{\sigma }^{lm}{}_{jk}.
\end{array}
\end{equation}
In some cases this expression can be further simplified with the help
of the characteristic equation of $\hat{\sigma }$.

\subsection{Summary of Relations in the Cartan Calculus} \label{S:SoRCC}

\paragraph{Commutation Relations}

For any $p$-form $\alpha $:
\begin{eqnarray}
\dl \alpha   & = & \dl(\alpha ) + (-1)^{p} \alpha  \dl\\
\Ix{i} \alpha  & = & \Ix{i}(\alpha ) + (-1)^{p}\Li_{O_{i}{}^{j}}(\alpha )
\Ix{j}\\
\Lix{i} \alpha  & = & \Lix{i}(\alpha )+\Li_{O_{i}{}^{j}}(\alpha )
\Lix{j} \label{il1}
\end{eqnarray}

\paragraph{Actions}

For any function $f \in \A$, 1-form $\omega _{f} \equiv Sf_{(1)} \dl
f_{(2)}$ and vector field $\phi \in \smash$:
\begin{eqnarray}
\Ix{i}(f) & = & 0\\
\Ix{i}(\dl f) & = & \dl f_{(1)}\inprod{\chi _{i}}{f_{(2)}}\\
\Ix{i}(\omega _{f}) & = & -\inprod{\chi _{i}}{Sf}\\
\Lix{i}(f) & = & \chi_i (f) = f_{(1)}\inprod{\chi_i}{f_{(2)}}\\
\Lix{i}(\omega _{f})& = & \omega _{f_{(2)}}\inprod{\chi_i}{S(f_{(1)})
f_{(3)}}\\
\Lix{i}(\phi )& = & \chi _{i(1)} \phi  S(\chi _{i(2)}) \label{il2}
\end{eqnarray}

\paragraph{Graded Quantum Lie Algebra of the Cartan Generators}

\begin{eqnarray}
\dl \dl & = & 0\\
\dl \Li_x & = & \Li_x \dl\\
\Lix{i} & = & \dl \Ix{i} + \Ix{i} \dl\\
\left[\Lix{i},\Lix{k} \right]_{q} & = & \Lix{l} f_{i}{}^{l}{}_{k}\\
\left[\Lix{i},\Ix{k} \right]_{q} & = & \Ix{l} f_{i}{}^{l}{}_{k}
\end{eqnarray}
The quantum commutator $[\, , \,]_{q}$ is here defined as follows:
\begin{equation}
\left[\Lix{i}, \Box \right]_{q} :=
\Lix{i} \Box - \Lio{i}{j}(\Box) \Lix{j}.
\end{equation}
This quantum Lie algebra becomes infinite-dimensional as soon as we
introduce derivatives along general vector fields.

\subsection{Universal Cartan Calculus}

In the case of a Universal Calculus (see section~\ref{S:UC}) the
relations of a Cartan Calculus can be expressed in a basis-free form
in Hopf algebra language. Here is a summary of commutation relations
valid on any form. All of these equations are identical to the
corresponding quantum Lie algebra relations when written in terms of
the bases $\{e_{\beta }\}$ and $\{f^{\beta }\}$, where $\beta =
0,1,\ldots,\infty$.  $x,y \in \U$, $a \in \A$, $\al$ is a $p$-form and
$\phi \in \smash$ is a vector field.
\begin{eqnarray}
\Li_{x} a & = & a_{(1)} \inprod{ x_{(1)}}{a_{(2)}} \Li_{x_{(2)}}\\
\Li_{x} \dg(a) & = & \dg(a_{(1)}) \inprod{ x_{(1)}}{a_{(2)}}
\Li_{x_{(2)}}\\
\Li_{x} \al & = & \Li_{x_{(1)}}(\al)\:\Li_{x_{(2)}}\\
\I_{x} a & = & a_{(1)} \inprod{x_{(1)}}{a_{(2)}} \I_{x_{(2)}}\\
\I_{x} \dg(a) & = & a_{(1)} \inprod{x-1 \epsilon (x)}{a_{(2)}}-
\dg(a_{(1)}) \inprod{x_{(1)}}{a_{(2)}} \I_{x_{(2)}}\\
\I_{x} \al & = & \I_{x}(\al) +(-1)^{p}
\Li_{x_{(1)}}(\al)\;\I_{x_{(2)}}\\
\dg \al & = & \dg(\al) + (-1)^{p} \al \dg\\
\dg \dg(\al) & = & - (-1)^{p} \dg(\al) \dg\\
\nonumber \\
\Li_{x}(\phi ) & = & x_{(1)} \phi S(x_{(2)}) \label{lixv}\\
\nonumber \\
\dg^{2}&=&0\\
\dg \Li_{x} &=& \Li_{x} \dg\\
\Li_{x} &=& \dg \I_{x} + 1 \epsilon (x) + \I_{x} \dg\z\mbox{\it
(universal Cartan identity)}\\
\Li_{x} \Li_{y} &=& \Li_{y^{(1)}} \inprod{x_{(1)}}{y^{(2)'}}
\Li_{x_{(2)}}\\
\Li_{x} \I_{y} &=& \I_{y^{(1)}} \inprod{x_{(1)}}{y^{(2)'}}
\Li_{x_{(2)}}
\end{eqnarray}
This type of Cartan calculus on an arbitrary Hopf algebra will be
treated in detail in an upcoming paper \cite{SchWatts}.

\subsection{Lie Derivatives Along General Vector Fields}

So far we have focused on Lie derivatives and inner derivations along
{\em left-invariant} vector fields, {\em i.e.} along elements of
$\tq$. The classical theory allows functional coefficients, so that a
general vector field need not be left-invariant.  Here we may
introduce derivatives along elements in the $\A\rtimes\tq$ plane by the
following set of equations valid on forms (recall $\epsilon (\chi ) =
0$ for $\chi \in {\cal T}_{q}$):
\begin{eqnarray}
\I_{f \chi_i } & = & f \Ix{i},\\
\Li_{f \chi_i } & = & \dl  \I_{f \chi_i } + \I_{f \chi_i } \dl,\\
\Li_{f \chi_i } & = & f \Lix{i} + \dl(f) \Ix{i}
\label{liefx},\\
\Li_{f \chi_i } \dl & = & \dl \Li_{f \chi_i }.
\end{eqnarray}
Equation (\ref{liefx}) can be used to define Lie derivatives
recursively on any form.

\section*{Acknowledgements}
We would like to thank P. Aschieri, S. Majid and N. Yu.  Reshetikhin
for helpful discussions.

This work was supported in part by the Director, Office of Energy
Research, Office of High Energy and Nuclear Physics, Division of High
Energy Physics of the U.S. Department of Energy under Contract
DE-AC03-76SF00098 and in part by the National Science Foundation under
grants PHY-90-21139 and PHY-89-04035.


\begin{thebibliography}{199}
\bibitem{Connes} A. Connes, {\it Publ. Math. IHES} {\bf 62} 257 (1985)
\bibitem{SWZ1} P. Schupp, P. Watts and B. Zumino, {\it Commun. Math.
Phys.} {\bf 157} 305 (1993)
\bibitem{RS} N. Yu. Reshetikhin and M. A. Semenov-Tian-Shansky, {\it
Lett. Math. Phys.} {\bf 19} (1990)
\bibitem{Jur} B. Jur\v{c}o, {\it Lett. Math. Phys.} {\bf 22} 177 (1991)
\bibitem{Maj} S. Majid, private communication (1993)
\bibitem{Zum} B. Zumino, in X$^{th}$ IAMP Conf. Leipzig (1991), ed. K.
Schm\"udgen, Springer-Verlag (1992)
\bibitem{SWZ2} P. Schupp, P. Watts and B. Zumino, {\it Lett. Math. Phys.}
{\bf 25} 139 (1992)
\bibitem{Drin} V. G. Drinfel'd, {\it Leningrad Math. J.} {\bf 1} 321
(1989)
\bibitem{Watts} P. Watts, ``Killing Form on Quasitriangular Hopf
Algebras and Quantum Lie Algebras'', in preparation
\bibitem{Wor} S. L. Woronowicz, {\it Commun. Math. Phys.} {\bf 122}
125 (1989)
\bibitem{CaMo} L. Castellani and A. R-Monteiro, preprint DFTT-18/93 (1993)
\bibitem{SchWatts} P. Schupp and P. Watts, ``Universal Cartan Calculus on
Hopf Algebras'', in preparation
\end{thebibliography}
\end{document}